\documentclass[preprint,a4paper,superscriptaddress,longbibliography]{revtex4-1}
\usepackage{graphicx,color,epsfig,subfig,dcolumn,bm,mathrsfs,amsmath}
\usepackage[T1]{fontenc}

\newcommand{\ud}{\mathrm{d}}

\DeclareMathAlphabet{\mathsfit}{\encodingdefault}{\sfdefault}{m}{sl}
\SetMathAlphabet{\mathsfit}{bold}{\encodingdefault}{\sfdefault}{bx}{sl}
\newcommand{\tens}[1]{\bm{\mathsfit{#1}}}

\newcommand{\change}[1]{\textcolor{black}{#1}}

\begin{document}


\title{Dynamics of a viscoelastic liquid filament connected to two mobile droplets}

\author{Jiajia Zhou}
\email[]{jjzhou@buaa.edu.cn}
\affiliation{Key Laboratory of Bio-Inspired Smart Interfacial Science and Technology of Ministry of Education, School of Chemistry, Beihang University, Beijing 100191, China} 
\affiliation{Center of Soft Matter Physics and Its Applications, Beihang University, Beijing 100191, China}
\author{Masao Doi}
\email[]{masao.doi@buaa.edu.cn}
\affiliation{Center of Soft Matter Physics and Its Applications, Beihang University, Beijing 100191, China}


\begin{abstract}

\change{A filament of liquid is usually unstable and breaks up into small droplets, while a filament of polymer solution is known to be quite stable against such instability, and  they form a stable configuration of filament connecting two spherical droplets.}
If the droplets are fixed in space, the liquid flows from the filament region to the droplet region to reduce the surface energy and the filament gets thinner.  
If the whole liquid is placed in another viscous fluid, the droplets approach each other, and the filament can get thicker. 
Here we study the dynamics of such system.  
We derive time evolution equations for the radius and the length of the filament taking into account the fluid flux from filament to droplets and the motion of the droplets.  
We will show that (a) if the centers of the droplets are fixed, the filament thins following 
the classical prediction of Entov and Hinch, and that (b) if the droplets are mobile (subject
to the Stokes drag in the viscous medium), the thinning of the filament is suppressed,
and under certain conditions, the filament thickens.  
This theory explains the phenomena observed by Yang and Xu \change{[Phys. Fluids {\bf 20}, 043101 (2008)]} in four-roller mill device. 
\end{abstract}


\maketitle


\section{Introduction}

A filament of Newtonian fluid is unstable due to Plateau-Rayleigh instability and breaks up into many small droplets to minimize the surface energy \cite{dBQ, Eggers1997}.
On the other hand, a filament of viscoelastic filament, formed by addition of small amount of long-chain polymers, is quite stable.  
Such phenomena have been used in rheometers.  
For example, in the Filament Stretching Rheometer (FSR), a fluid sample is stretched at a controlled force or rate \cite{McKinley2002, Bach2003a},  and the filament diameter is measured to study the extensional properties.
In another apparatus called Capillary Breakup Extension Rheometer (CaBER), two plates containing a viscoelastic fluid are separated rapidly for a period of time and kept stationary afterward \cite{Anna2001}. 
The stretched liquid then quickly takes the configuration of liquid filament connecting two droplets attached to the plates, and \change{some} rheological data is obtained from the thinning and break-up behavior of the filament \change{\cite{Dinic2019, Dinic2019a}}.

In these setups, the position of the droplets is controlled externally. 
A different situation has been studied by Yang and Xu \cite{YangJianmao2008}.  
They placed a droplet of polymer solution in a four-roll mill rheometer cell \cite{Taylor1934, Fuller1980}, and repeated the operation of stretching and relaxation of the cell.  
This brings the polymer solution into the configuration shown in Fig.~\ref{fig:sketch}, where two droplets are connected by a filament. 
They observed the relaxation of this configuration when the external flow is stopped.  
The relaxation is caused by two forces, one is the capillary force which tends to reduce the surface area of the filament, and the other is the relaxation of the stretched polymer in the filament.
The length of the filament decreases in time, but the diameter of the filament increases, or remains constant in time. 

In this paper, we will conduct a theoretical analysis for their experiment.  
Although the phenomena is a simple relaxation of a stretched filament, theoretical analysis is not simple since there is a transport of material between the filament region to the droplet region.  
To calculate the time evolution of the system,  one has to calculate the fluid flux flowing from filament to droplets.  
This problem has the same difficulty as the problem of CaBER experiments. 

The theory of the filament relaxation in CaBER, \emph{i.e.}, the relaxation of a filament connected to fixed droplets, was first discussed by Entov and Hinch \cite{Entov1997}. 
Their theory successfully explained the main feature of the experiments, but it included an assumption on the flux at the filament/droplet interface.  
This assumption was questioned in subsequent works, and many works have been done to resolve the issue \cite{Clasen2006, Bazilevskii2014, Bazilevskii2015, 2018_ve_filament}.

In our previous work \cite{2018_ve_filament}, we have derived a new assumption to solve the CaBER problem using the Onsager principle \cite{DoiSoft}, and shown that it gives results consistent with numerical solution of Clasen \emph{et al.} \cite{Clasen2006}.
\change{Extending our previous work \cite{2018_ve_filament} we here consider the case} that the beads can move in a viscous medium, and take into account of two new sources of energy dissipation: 
One is the dissipation due to the motion of the droplets, and the other is the energy dissipation associated with the flux at the transition region between the filament and droplets.
We shall analyze the time dependence of the filament radius and
show that it qualitatively explain the experiments of Yang and Xu \cite{YangJianmao2008}.

This article is organized as follows: 
In Sec.~\ref{sec:model}, we introduce our model for the viscoelastic filament and derive 
general governing equations for the system shown in Fig.~\ref{fig:sketch}. 
In Sec.~\ref{sec:CaBER}, we study the CaBER case and discuss how the extra energy dissipation at the filament/droplet interface affects our previous analysis.  
In Sec.~\ref{sec:BSB}, we study the case of mobile droplets, and compare the theoretical results with the experimental results.
We conclude with a brief summary in Sec.~\ref{sec:summary}.

\section{Theoretical formulation}
\label{sec:model}

\subsection{Description of the problem}
\begin{figure}[htbp]
  \centering
  \includegraphics[width=0.5\textwidth]{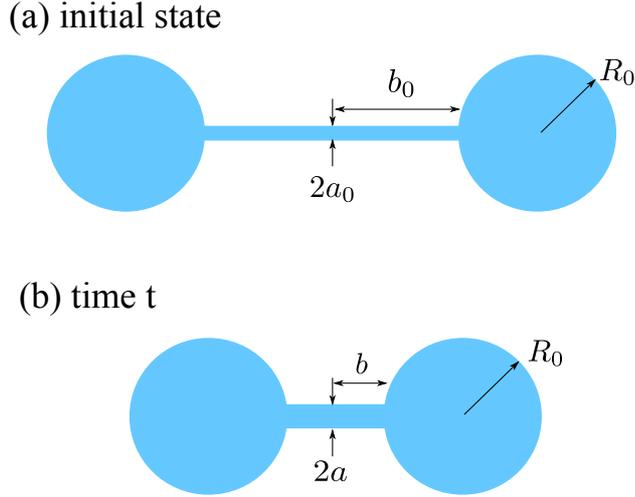}
  \caption{Schematic pictures of the time evolution of the system. \change{(a) The initial state at $t=0$. The filament has a diameter of $2a_0$ and a length of $2b_0$. The radius of the bead is $R_0$. (b) At later time $t$, the diameter and the length of the filament change to $2a$ and $2b$, respectively. The increase of the bead radius is neglected. }}
  \label{fig:sketch}
\end{figure}

We consider the system shown schematically in Fig.~\ref{fig:sketch}. 
Initially,  the polymer solution takes the configuration shown in Fig.~\ref{fig:sketch}(a), where
two spherical droplets (which will be referred to as beads in the following)
of radius $R_0$ are connected by a thin cylindrical filament (which will be 
referred to as string)  of radius $a_0$ and length $2b_0$.  
The polymer solution is placed in a medium Newtonian liquid of viscosity $\eta_m$.  

\vspace{0.3cm}

As time goes on, the radius and the length of the filament change in time, due to the interfacial tension $\gamma$ of the filament and due to the stretched polymers in the filament.  
We shall focus on the time dependence of the radius $a(t)$ and the length $b(t)$ of the filament.
The radius of the  droplet $R(t)$ also changes in time, but since this change is very small, we assume $R(t)=R_0$ for all time.

The volume of the filament at time $t$ is given by $V(t)=2 \pi a^2(t) b(t)$, shown in Fig.~\ref{fig:sketch}(b). 
This is different from the initial volume $ V_0 = 2 \pi a_0^2 b_0$ since the polymer solution can move from the filament region to the bead region. 
\change{The resulting increase of the bead radius is neglected because $a_0 \ll R_0$ while $b_0$ and $R_0$ are of the same order.}
Let $x(t)$ be the filament volume normalized by $V_0$,
\begin{equation}
  V(t) = x(t) V_0 .
\end{equation}
Let $\lambda(t)$ be the stretching ratio of the filament, which is defined by the change of the filament radius
\begin{equation}
  \lambda(t)=\left( \frac{a_0}{a(t) } \right )^2 .
\end{equation}
The length of the filament is then written as
\begin{equation}
 b(t)=\lambda(t) x(t) b_0.
\end{equation}

To describe the viscoelasticity of the solution, we use the Oldroyd-B model.  
The flow in the present problem is uniaxial extension, and the strain rate is given by
$\dot \epsilon=\dot \lambda/\lambda$.  
The Oldroyd-B model gives the following expression for the the extensional stress $\sigma_{zz}- \sigma_{rr}$ ($z$ and $r$ being the axial and the radial directions of the filament).
\begin{equation}
  \sigma_{zz} - \sigma_{rr} = 3 \eta_s \dot \epsilon  + G( c_{zz}- c_{rr})
\end{equation}
where $\eta_s$ is the viscosity of the solvent, $G$ is the elastic modulus of the polymer and
$c_{zz}$ and $c_{rr}$ are the $zz$ and $rr$ component of the conformation tensor $\tens{c}$ (a tensor representing the conformation of the polymer in the solution).
The constitutive equation of the Oldroyd-B model is then written as
\begin{eqnarray}
  \dot c_{zz} &=&  2\dot \epsilon c_{zz} - \frac{1}{\tau}( c_{zz} -1)  , \\
  \dot c_{rr} &=&  - \dot \epsilon c_{rr} - \frac{1}{\tau}( c_{rr} -1) ,
\end{eqnarray}
where $\tau$ is the relaxation time of the polymer.  
Given the constitutive equation, the problem can be solved by continuum mechanics.
This was indeed done by Clasen \emph{et al.} \cite{Clasen2006}: assuming the cylindrical symmetry of
the system, they set up the evolution equation for the filament shape and solved it numerically.
Here we use a different approach, and this is detailed in the following section.

\subsection{Onsager principle}

In our previous paper \cite{2018_ve_filament}, we have shown that the set of equations for Oldroyd-B fluid model in the inertialess system can be derived from the Onsager principle, a general principle describing the time evolution of many non-equilibrium systems.  
The principle is based on the observation that many time evolution equations in non-equilibrium systems can be written in the following form
\change{\begin{equation}
  \sum_j \zeta_{ij} \dot \phi_j + \frac{\partial A}{\partial \phi_i} =0   
  \label{eqn:O1}    
\end{equation}}
where \change{$\phi = (\phi_1, \phi_2, \cdots)$} stands for the set of variables specifying the non-equilibrium state, $\zeta_{ij}$ is the friction matrix, and \change{$A(\phi)$} is the free energy. Equation (\ref{eqn:O1}) represents the balance of two forces, one is the potential force which tends to bring the system to the state of free energy minimum, and the other is the frictional force which resists this change.  
Due to the reciprocal relation $\zeta_{ij}=\zeta_{ji}$, Eq.~(\ref{eqn:O1}) can be written as 
\change{\begin{equation}
  \frac{\partial }{\partial \dot \phi_i} \left[ \Phi(\dot \phi) 
    + \dot A(\dot \phi) \right] =0 ,
  \label{eqn:O2}    
\end{equation}}
where \change{$\Phi(\dot \phi)$} and \change{$\dot A(\dot \phi)$} are functions of \change{$ \dot \phi = (\dot \phi_1, \dot  \phi_2,...)$}, defined by
\change{\begin{equation}
  \Phi(\dot \phi)= \frac{1}{2}\sum_{ij} \zeta_{ij} \dot \phi_i \dot \phi_j, \qquad
  \dot A (\dot \phi) = \sum_i \frac{\partial A }{\partial \phi_i} \dot \phi_i .
  \label{eqn:O3}    
\end{equation}}
Equation (\ref{eqn:O2}) indicates that the time evolution of the system is determined by the
condition that the function \change{$ \Phi(\dot \phi) + \dot A(\dot \phi )$} is the minimum with respect to \change{$\dot \phi$}.  
The function \change{$ \Phi(\dot \phi)$} is called energy dissipation function and represents the energy dissipation rate when the system is evolving at rate \change{$\dot \phi$}.
The function \change{$\dot A(\dot \phi)$} represents the change rate of free energy, and the function \change{$\mathscr{R}(\dot \phi)=\Phi(\dot \phi) + \dot A(\dot \phi )$} is called Rayleighian.
\change{The Onsager principle has been applied to various problems in fluid mechanics \cite{XuXianmin2016, DiYana2016, 2018_coating, 2018_square, 2019_Taylor_rising}.}

In the present problem, the nonequilibrium state of the system is characterized by variables $a(t)$, $b(t)$, $c_{zz}(t)$, and $c_{rr}(t)$. 
To simplify the analysis, we use the same approximation as in our previous paper \cite{2018_ve_filament} and assume that $c_{zz}$ and $c_{rr}$ are characterized by a single parameter $\lambda_p(t)$,
\begin{equation}
  c_{zz}(t) =  \lambda_p^2(t), \quad c_{rr}(t) = \frac{1}{\lambda_p(t)} .
\end{equation}
Here $\lambda_p(t)$ represents the stretching of the polymer chain in the filament. 

In the following we use three dimensionless variables to characterize the time evolution, $\lambda(t)=[a_0/a(t)]^{1/2}$, $x(t)=a^2(t)b(t)/(a_{0}^{2}b_0)$ and $\lambda_p(t)$.  
We shall construct the Rayleighian as a function $\dot \lambda$,  $\dot x$, and $\dot \lambda_p$.

\subsection{Free energy}

The free energy of the system is given by
\begin{equation}
  \label{eq:freeE}
  A =   \frac{G}{2} 2\pi a^2 b  \left(\lambda_p^2 + \frac{2}{\lambda_p} \right) 
    + 4 \pi ab \gamma 
   = x G V_0 \left[ \frac{1}{2} \left(\lambda_p^2 + \frac{2}{\lambda_p} \right) 
    + 2 \alpha \sqrt{\lambda} \right].    
\end{equation}
The first term is the elastic energy due to the stretching of the polymer chains, and the second term is the contribution from the interfacial energy.  
Here $\alpha$ is the elasto-capillary number defined by
\begin{equation}
  \alpha = \frac{\gamma}{a_0 G}.
\end{equation}
The free energy contribution from the beads is neglected as we are assuming that the polymer chains in the beads are relaxed.

From Eq.~(\ref{eq:freeE}), $\dot A$ is calculated as
\begin{equation}
  \label{eq:Adot}
  \dot{A} = G V_0 \left[ \dot{x} \left( \frac{\lambda_p^2}{2} + \frac{1}{\lambda_p} 
    + 2 \alpha \sqrt{\lambda} \right) 
    + x \left( \lambda_p \dot{\lambda}_p - \frac{1}
      {\lambda_p^2} \dot{\lambda}_p + \alpha \lambda^{-1/2} \dot{\lambda} \right) \right] .
\end{equation}

\subsection{Dissipation function}

The dissipation function has three contributions
\begin{equation}
     \Phi= \Phi_{\rm string} + \Phi_{\rm bead } + \Phi_{\rm trans} .
\end{equation}

$\Phi_{\rm string}$ represents the energy dissipation taking place in the string part, and is the same as in our previous work \cite{2018_ve_filament}
\begin{equation}
  \Phi_{\rm string} = x V_0 \left[ \frac{3}{2} \eta_s 
    \left( \frac{\dot{\lambda}}{\lambda} \right)^2 
    + \eta_p \left( \lambda_p^2 + \frac{1}{2 \lambda_p} \right) 
    \left(  \frac{\dot{\lambda_p}}{\lambda_p} - \frac{\dot{\lambda}}{\lambda} \right)^2 
    \right] . 
  \label{eq:Phi_string}
\end{equation}
The two terms in Eq.~(\ref{eq:Phi_string}) \change{respectively account for the solvent and polymer contribution to the dissipation in the uniaxial flow in the string.}

$\Phi_{\rm bead}$ represents the energy dissipation caused by the motion of the beads in the medium fluid
\begin{equation}
  \Phi_{\rm bead} = 2 \times \frac{1}{2} \zeta_{\rm bead} \dot{b}^2 
  = \zeta_{\rm bead} b_0^2 ( \dot{\lambda} x + \lambda \dot{x} )^2,
  \label{eq:Phi_bead}
\end{equation}
where $\zeta_{\rm bead}$ represents the friction constant of the bead. 
If the beads behave as a solid, the friction constant is given by $\zeta_{\rm bead} = 6\pi \eta_m R_0$. 

$\Phi_{\rm trans}$ represents the energy dissipation taking place in the transition region connecting the string and the bead  \cite{Bazilevskii2014, Bazilevskii2015}.
The polymer solution enters the bead with the velocity
\begin{equation}
  v= \frac{ \dot{V} }{ 2 \pi a^2 } 
  = \frac{ V_0 \dot{x}}{2 \pi a_0^2 / \lambda} 
  = b_0 \lambda \dot{x} .
\end{equation}
This velocity quickly decays to zero in the bead region following the equation
\begin{equation}
  v_r = \frac{a^2}{2r^2} v,  \quad r>a . 
  \label{eqn:D1}
\end{equation}
The energy dissipation caused by such flow is proportional to $v^2$, and can be written as
\begin{equation}
  \Phi_{\rm trans}   = 2 \times \frac{1}{2}\zeta_{\rm t}v^2 =\zeta_{\rm t}v^2 .
  \label{eq:Phi_t1}
\end{equation}
This defines the friction constant $\zeta_{\rm t}$ which represents the energy dissipation
taking place in the transition region. 
Precise calculation of $\zeta_{\rm t}$ is difficult since it involves the transient flow of viscoelastic fluid, but $\zeta_{\rm t}$ can be estimated by the following argument.  

In the transition region, the solvent exerts no frictional force on the polymer as it moves
with the same velocity as polymer. Hence the main part of the energy dissipation associated with
the relative motion between polymer and solvent [\emph{i.e.}, the second term 
$(\dot \lambda/\lambda - \dot \lambda_p/\lambda_p)^2$ in Eq.~(\ref{eq:Phi_string})] is ignorable. 
Therefore the energy dissipation can be estimated by the 
dissipation of Newtonian fluid of viscosity $\eta_s$ \cite{Landau07}
\begin{eqnarray}
  \Phi_{\rm trans} &=& 2 \times \frac{1}{4} \eta_s 
                         \int_{|\mathbf{r}|>a } \ud \mathbf{r} \, 
    \left( \frac{\partial v_{\alpha}}{\partial r_{\beta}} 
    + \frac{\partial v_{\beta}}{\partial r_{\alpha}} \right)^2 \nonumber \\
  &=& \frac{1}{2} \eta_s \int_a^{\infty} \ud r \, 2\pi r^2 
  \, 4 \left[  \left( \frac{\partial v_r}{\partial r} \right)^2 
    + 2\left( \frac{v_r}{r} \right)^2  \right] .
      \label{eq:Phi_t}
\end{eqnarray}
From Eqs.~(\ref{eq:Phi_t}) and (\ref{eqn:D1}), we have
\begin{equation}
  \Phi_{\rm trans} 
     =   2 \pi \eta_s a v^2 .
     \label{eq:Phi_t2}
\end{equation}
Comparing Eq.~(\ref{eq:Phi_t2}) with Eq.~(\ref{eq:Phi_t1}), we have
\begin{equation}
  \zeta_{\rm t}  \simeq 2 \pi \eta_s a  =  2\pi \eta_s a_0 \lambda^{-1/2}   .
  \label{eq:Phi_t3}
\end{equation}

Hence the dissipation function of the whole system is given by
\begin{equation}
  \Phi = \eta_p V_0 \Bigg[ x \left( \frac{3}{2} \frac{\eta_s}{\eta_p} 
         \left( \frac{ \dot{\lambda}}{\lambda} \right)^2 
        + \left( \lambda_p^2 + \frac{1}{2}\lambda_p \right) 
          \left( \frac{ \dot{\lambda}_p}{\lambda_p} - \frac{ \dot{\lambda}}{\lambda} \right)^2 \right) 
        + \frac{1}{2} \beta_{\rm b} (\dot{\lambda} x + \lambda \dot{x})^2 
        + \frac{1}{2} \beta_{\rm t} \lambda^{3/2} \dot{x}^2 \Bigg]
\end{equation}
where the two coefficients are given by
\begin{eqnarray}
  \label{eq:betab}
  \beta_{\rm b} &=& \frac{ 12 \pi \eta_m R_0 b_0^2 }{\eta_p V_0} = 6 \frac{\eta_m}{\eta_p} \frac{R_0 b_0}{a_0^2}, \\
  \label{eq:betat}
  \beta_{\rm t} &=& \frac{ 4 \pi \eta_s a_0 b_0^2 }{\eta_p V_0} = 2 \frac{\eta_s}{\eta_p} \frac{b_0}{a_0}.
\end{eqnarray}

\subsection{Evolution equations}

The evolution equations are obtained by minimizing Rayleighian 
$\mathscr{R} = \Phi + \dot A$
with respect to $\dot \lambda$, $\dot \lambda_p$ and $\dot x$ \cite{DoiSoft}:
\begin{eqnarray}
  \label{eq:dlam0}
  \frac{\partial \mathscr{R}}{\partial \dot{\lambda}} = 0 & \quad \Rightarrow \quad&
    3 \frac{\eta_s}{\eta_p} \frac{\dot{\lambda}}{\lambda} + \frac{1}{\tau} ( \lambda_p^2 - \frac{1}{\lambda_p} ) + \beta_{\rm b} \lambda (\dot{\lambda} x + \lambda \dot{x}) + \frac{\alpha}{\tau} \sqrt{\lambda} =0 , \\
  \label{eq:dlamp}
  \frac{\partial \mathscr{R}}{\partial \dot{\lambda}_p} = 0 & \quad \Rightarrow \quad&
    \frac{ \dot{\lambda}_p}{\lambda_p} - \frac{\dot{\lambda}}{\lambda}
 = - \frac{1}{2\tau} \frac{ \lambda_p^2 - \frac{1}{\lambda_p}} {\lambda_p^2 + \frac{1}{2\lambda_p}} , \\
  \label{eq:dx}
  \frac{\partial \mathscr{R}}{\partial \dot{x}} = 0 & \quad \Rightarrow \quad&
    \beta_{\rm b} \lambda (\dot{\lambda} x + \lambda \dot{x}) + \beta_{\rm t} \lambda^{3/2} \dot{x} + \frac{1}{\tau} \left( \frac{1}{2} \lambda_p^2 + \frac{1}{\lambda_p} + 2 \alpha \sqrt{\lambda} \right) = 0 .
\end{eqnarray}

Equation (\ref{eq:dlam0}) represents the force balance exerted on the string. 
The first two terms account the tensile force in the filament and the third term is the friction force exerted by the beads, and the last term is the contribution from the surface tension.
Equation (\ref{eq:dlamp}) represents the evolution equation for the polymer conformation $\lambda_p$, which is essentially equivalent to the constitutive equation of the polymer solution.
Equation (\ref{eq:dx}) represents the coexistence condition between the filament and the bead \cite{2018_ve_filament}.

It is convenient to introduce another state variable $y=\lambda x$, which represents the filament length $b=y b_0$.
Using the relations
\begin{equation}
  \label{eq:doty}
  \dot{y} = \dot{\lambda} x + \lambda \dot{x} , \quad
  \dot{x} = \frac{\dot{y} \lambda - y \dot{\lambda}}{\lambda^2} = \frac{y}{\lambda} \left( \frac{ \dot{y}}{y} - \frac{\dot{\lambda}}{\lambda} \right),
\end{equation}
we can rewrite Eqs.~(\ref{eq:dlam0}) and (\ref{eq:dx}) as
\begin{eqnarray}
  \label{eq:dlam}
  3 \frac{\eta_s}{\eta_p} \frac{\dot{\lambda}}{\lambda} 
     + \beta_{\rm b} \lambda \dot{y} 
  &=& - \frac{1}{\tau} \left( \lambda_p^2 - \frac{1}{\lambda_p} + \alpha \sqrt{\lambda} \right) , \\
  \label{eq:dy}
   \beta_{\rm t} y \frac{ \dot{\lambda} }{\sqrt{\lambda}}
     - ( \beta_{\rm b} \lambda + \beta_{\rm t} \sqrt{\lambda} ) \dot{y} 
  &=& \frac{1}{\tau} \left( \frac{1}{2} \lambda_p^2 + \frac{1}{\lambda_p} + 2 \alpha \sqrt{\lambda} \right)  .
\end{eqnarray}

In order to conduct analytical calculation, we make further simplification: 
(a) Since  $\lambda_p \gg 1$  (polymers are strongly stretched) in the following analysis,  
we neglect the terms of $O(1/\lambda_p)$ on the right hand side of 
Eqs.~(\ref{eq:dlamp}), (\ref{eq:dlam}) and (\ref{eq:dy}).
(b) Since $\eta_p \gg \eta_s$, we ignore the term involving $\eta_s / \eta_p$ in Eq.~(\ref{eq:dlam}).

With these approximations, the evolution equations (\ref{eq:dlamp}), (\ref{eq:dlam}) and (\ref{eq:dy}) 
are simplified as
\begin{eqnarray}
  \label{eq:dlamp2}
  \frac{ \dot{\lambda}_p}{\lambda_p} - \frac{\dot{\lambda}}{\lambda}
     &=& - \frac{1}{2\tau} ,\\
    \label{eq:dy2}
  \dot{y} &=& - \frac{1}{\tau \beta_{\rm b} \lambda } 
       \left( \lambda_p^2  + \alpha \sqrt{\lambda} \right), \\
  \label{eq:dlam2}
  \dot{\lambda} &=& - \frac{\sqrt{\lambda}}{\tau \beta_{\rm t} y } 
     \left(\frac{1}{2} \lambda_p^2 - \alpha \sqrt{\lambda} \right) 
     - \frac{1}{ \tau \beta_{\rm b} } 
       \left( \lambda_p^2 + \alpha \sqrt{\lambda} \right) .
\end{eqnarray}  

The initial condition is
\begin{equation}
  \label{eq:ic}
  \lambda(t)=1, \quad \lambda_p(0)=\lambda_{p0}, \quad y(0)=1,
\end{equation}
where $\lambda_{p0}$ is the initial value of $\lambda_p$.  
The coupled differential equations (\ref{eq:dlamp}), (\ref{eq:dlam}), and (\ref{eq:dy}) can be solved numerically with the initial conditions (\ref{eq:ic}). 

There are four input parameters that determine the time evolution of the system:
\begin{itemize}
\item $\alpha = {\gamma}/(a_0 G)$, the elasto-capillary number that characterizes 
the importance of the surface tension relative to the elastic force of polymer.
\item $\beta_{\rm t}$, the dissipation coefficient in the transition region.
\item $\beta_{\rm b}$, the dissipation coefficient for the movement of the two large droplets.
\item $\lambda_{p0}$, the stretching of the polymer chain at $t=0$. 
\end{itemize}

The CaBER case corresponds to the limit $\beta_{\rm b} \rightarrow \infty$.
In the following, we first consider this case and study the effect of the transition dissipation on the thinning of liquid filament.
Next we consider the case of finite $\beta_{\rm b}$, and study the effect of the ratio $\beta_{\rm t}/\beta_{\rm b}$  on the dynamics of liquid filament.

\section{CaBER CASE}
\label{sec:CaBER}


When $\beta_{\rm b} \rightarrow \infty$, the beads cannot move. 
This corresponds to CaBER experiments. 
In this case, we may set $y(t)=1$, and the time evolution equations are simplified as 
\begin{eqnarray}
  \label{eq:dlampA}
  \frac{ \dot{\lambda}_p}{\lambda_p} - \frac{\dot{\lambda}}{\lambda}
     &=& - \frac{1}{2\tau} ,\\
  \label{eq:dlamA}
  \dot{\lambda} &=& - \frac{\sqrt{\lambda}}{\tau \beta_{\rm t} } 
     \left(\frac{1}{2} \lambda_p^2 - \alpha \sqrt{\lambda} \right).
\end{eqnarray}

\begin{figure}[htbp]
  \centering
  \includegraphics[width=1.0\textwidth]{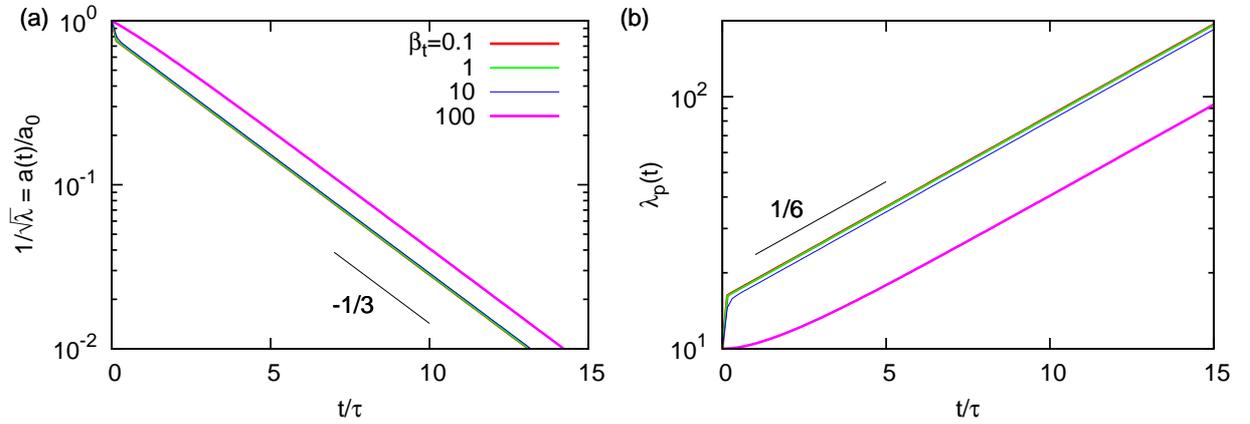}
  \caption{Evolution of the system in the limit of $\beta_{\rm b} \rightarrow \infty$. The parameters are $\alpha=100$ and $\lambda_{p0}=10$.  (a) The time evolution of the string radius $a(t)/a_0$, calculated from Eq.~(\ref{eq:solA_lam}). (b) The time evolution of the polymer stretching $\lambda_p(t)$ calculated from Eq.~(\ref{eq:solA_lamp}).}
  \label{fig:CaBER}
\end{figure}

Figure \ref{fig:CaBER} shows the result of the numerical solutions for 
$\alpha=100$ and $\beta_{\rm t}=0.1$, 1, 10 and 100.  
It is seen that for small $\beta_{\rm t}$ value, both $\lambda(t)$ and $\lambda_p(t)$ change rapidly in the initial stage, and then vary exponentially.  
Such behavior can be understood from the analytical solution of Eqs.~(\ref{eq:dlampA}) and
(\ref{eq:dlamA}).

Equation~(\ref{eq:dlampA}) can be integrated to give  
\begin{equation}
  \label{eq:solA_lamp}
  \lambda_p(t) = \lambda_{p0} \lambda(t) \exp \left(-\frac{t}{2\tau} \right).
\end{equation}
Substituting the above equation into Eq.~(\ref{eq:dlamA}), we get
\begin{equation}
  \label{eq:dlamA2}
  \beta_{\rm t} \tau \dot{\lambda} = - \left[ \frac{1}{2} \lambda_{p0}^2 \, \lambda^{5/2} 
    \exp \left( -\frac{t}{\tau} \right) - \alpha \lambda \right] .
\end{equation}
The solution to Eq.~(\ref{eq:dlamA2}) is 
\begin{equation}
  \label{eq:solA_lam}
  \lambda(t) = \left[ \frac{\lambda_{p0}^2}{ 2\alpha - \frac{4}{3} \beta_{\rm t} } 
    \left( \exp \left( - \frac{t}{\tau} \right) 
         - \exp \left( - \frac{3\alpha}{2 \beta_{\rm t}} \frac{t}{\tau} \right) \right) 
    + \exp \left( - \frac{3\alpha}{2\beta_{\rm t}} \frac{t}{\tau} \right) \right]^{-2/3} .
\end{equation}
Equations (\ref{eq:solA_lam}) indicates that the system has two characteristic relaxation times, $\tau$ and $\tau_s = \tau \beta_{\rm t}/\alpha$.  
In the case of $\tau_s \ll \tau$,  $\lambda(t)$ quickly changes from the initial value $\lambda(0)=1$ to the asymptotic exponential form
\begin{equation}
  \label{eq:CaBER_lam}
  \lambda(t) = \left( \frac{ 2\alpha - \frac{4}{3} \beta_{\rm t} }{\lambda_{p0}^2} \right)^{2/3}  
                         \exp \left( \frac{2t}{3\tau} \right) .
\end{equation}
The asymptotic form of $\lambda_p(t)$ is obtained from Eqs.~(\ref{eq:CaBER_lam}) and (\ref{eq:solA_lamp}),
\begin{equation}
  \label{eq:CaBER_lamp}
  \lambda_p = \lambda_{p0} \left( \frac{ 2\alpha - \frac{4}{3} \beta_{\rm t} }{\lambda_{p0}^2} \right)^{2/3}  \exp \left( \frac{t}{6\tau} \right).
\end{equation}
Hence both $\lambda(t)$ and  $\lambda_p(t)$  increases exponentially in time. 
Such exponential time-dependence is in agreement with the classical theory of Entov and Hinch \cite{Entov1997}.  
Here we emphasize a significance of the analytical solution (\ref{eq:solA_lam}) in the context of theoretical development for CaBER problem. 

The classical theory of Entov and Hinch included a disputable equation which they introduced to close their set of equations.
The extra equation represents a condition for the coexistence of bead and filament.  
This extra equation has been questioned in subsequent works.
Different assumptions have been proposed and studied \cite{Clasen2006, Bazilevskii2014, Bazilevskii2015, 2018_ve_filament}.
Surprisingly, such studies asymptotically gave the same exponential growth for $\lambda(t)$ and  
$\lambda_p(t)$.  
Different assumptions only gave different front factors for the exponential functions.  
Since the front factor can be absorbed in the uncertainty of the initial condition, this issue on the coexistence condition has not been pursued further, and the problem has remained unsolved.

The controversy on the extra condition in the Entov and Hinch theory has been resolved recently by Clasen \emph{et al.} \cite{Clasen2006}.  
They set up the equation for the shape of the viscoelastic filament connected to spherical droplets, and solved it numerically. 
They also have shown that the asymptotic solution satisfies the following condition
\begin{equation}
  \label{eq:CaBER_A1}  
  \lambda_p^2 =2  \alpha \sqrt{\lambda} .
\end{equation}
This condition is consistent with their numerical calculation.

In our previous paper \cite{ 2018_ve_filament}, we have derived Eq.~(\ref{eq:CaBER_A1}) from thermodynamic argument: Equation~(\ref{eq:CaBER_A1}) represents the condition that the chemical potential of the fluid in the filament is equal to that in the beads.  
This argument is valid when the energy dissipation taking place at the filament/bead interface is ignored.  
In fact, if take the limit of $\beta_t \to 0$ in Eq.~(\ref{eq:dlamA}), we get Eq.~(\ref{eq:CaBER_A1}).  

The analytical solution (\ref{eq:solA_lam}) gives a correction to the asymptotic solution.
Equation (\ref{eq:solA_lam}) indicates that the asymptotic solution is altered in two
respects:
\begin{itemize}
\item[(a)] The front factor of the exponential function for $\lambda(t)$ is altered.
This change is on the order of $\beta_{\rm t}/\alpha$, which is small due to the smallness of $\beta_{\rm t}$. 
The asymptotic relation between $\lambda_p$ and $\lambda$ is modified to 
\begin{equation}
  \lambda_p^2 = ( 2\alpha - \frac{4}{3} \beta_{\rm t} ) \sqrt{\lambda} .
\end{equation}
This is consistent with Eq.~(\ref{eq:CaBER_A1}) when taking the limit $\beta_{\rm t} \rightarrow 0$.
\item[(b)] The short time behavior of $\lambda(t)$ is altered.  
The asymptotic solution (\ref{eq:CaBER_lam}) does not satisfy the initial condition (\ref{eq:ic}), but this inconsistency is resolved by the analytical solution (\ref{eq:solA_lam}) which bridges the initial state to the asymptotic solution with the relaxation time $\tau_s=\tau \beta_{\rm t}/\alpha$.
\end{itemize}

\section{Bead-string-bead case}
\label{sec:BSB}

Now we study the case of finite $\beta_{\rm b}$, and discuss the  system studied by Yang and Xu \cite{YangJianmao2008}. 
The CaBER problem corresponds to the case of  $\beta_{\rm b} \gg \beta_{\rm t}$.
Here we first study the case of the other limit, $\beta_{\rm b} \ll \beta_{\rm t}$, and then discuss how the behavior changes as the ratio $\beta_{\rm b} /\beta_{\rm t}$ increases.

\subsection{$\beta_{\rm b}/ \beta_{\rm t} \ll 1$ case, elastic limit }
\label{sec:BSB1}

We first discuss the case of $\beta_{\rm b}/ \beta_{\rm t} \ll 1$. 
Since $\beta_{\rm b}$ is finite, the length of the filament $b(t)=y(t) b_0$ 
changes in time following Eq.~(\ref{eq:dy2})
\begin{equation}
  \dot{y} = - \frac{1}{\tau \beta_{\rm b} \lambda} 
    \left( \lambda_p^2 + \alpha \sqrt{\lambda} \right). 
\end{equation}
The first term in the brackets represents the elastic force of the polymer, and the second term represents the surface tension; both forces pull the beads inwardly, and the filament length $b(t)$ always decreases in time.
The retraction of the beads ends when the two beads touch each other, \emph{i.e.}, when $y$ becomes equal to zero.  
This time is of the order of  $\beta_{\rm b} \tau$, which is much shorter than $\tau$ for $\beta_{\rm b} \ll \beta_{\rm t} <1$. 
Therefore if the beads are mobile,  the dynamics takes place in a time scale much shorted than the viscoelastic relaxation time $\tau$. 
In such a short time, the viscoelastic filament behaves as an elastic filament.  
This is indeed seen from the previous equations.
In the time scale much smaller than $\tau$, Eq.~(\ref{eq:dlamp2}) is simplified as 
$\dot{\lambda}_p/\lambda_p - \dot{\lambda}/\lambda =0$, which indicates that 
the polymer in the filament changes in the same way as in a gel, i.e., changes affinely.

It must be noted that although the filament behaves as an elastic string, the relaxation dynamics of our system is different from the relaxation of an elastic filament in a viscous medium.  
This is because the volume of the filament changes in time in the present problem.  

The time evolution of $\lambda$ in the present problem is given by Eq.~(\ref{eq:dlam2}). 
In the limit of $\beta_{\rm b} \rightarrow 0$, this equation is simplified as 
\begin{equation}
  \label{eq:dlamB1}
  \dot{\lambda} = - \frac{1}{\tau \beta_{\rm b}} 
    \left( \lambda_p^2 + \alpha \sqrt{\lambda} \right).
\end{equation}
Since the right-hand-side is negative, $\lambda$ decreases with time and the radius of the string $a=a_0/\sqrt{\lambda}$ increases with time. 
This is similar to that of the elastic filament, but the dynamics of the present problem is more complicated.

For an elastic string, the volume is constant ($x=y/\lambda={\rm const.}$)
This leads to $\dot{y}/y=\dot{\lambda}/\lambda$. On the other hand, in 
the present problem,  we have, from Eqs.~(\ref{eq:dy2}) and (\ref{eq:dlamB1}),
\begin{equation}
  \label{eq:dyB1}
  \dot{y} = \frac{\dot{\lambda}}{\lambda} 
  \quad \Rightarrow \quad
  y = 1 + \ln \lambda.
\end{equation}
Thus $\dot{y}/y$ is not equal to $\dot{\lambda}/\lambda$.  
This is because during the relaxation the volume of the 
filament $x=y/\lambda=(1+ \ln \lambda)/\lambda$ decreases in time.

\begin{figure}[!htbp]
  \centering
  \includegraphics[width=1.0\textwidth]{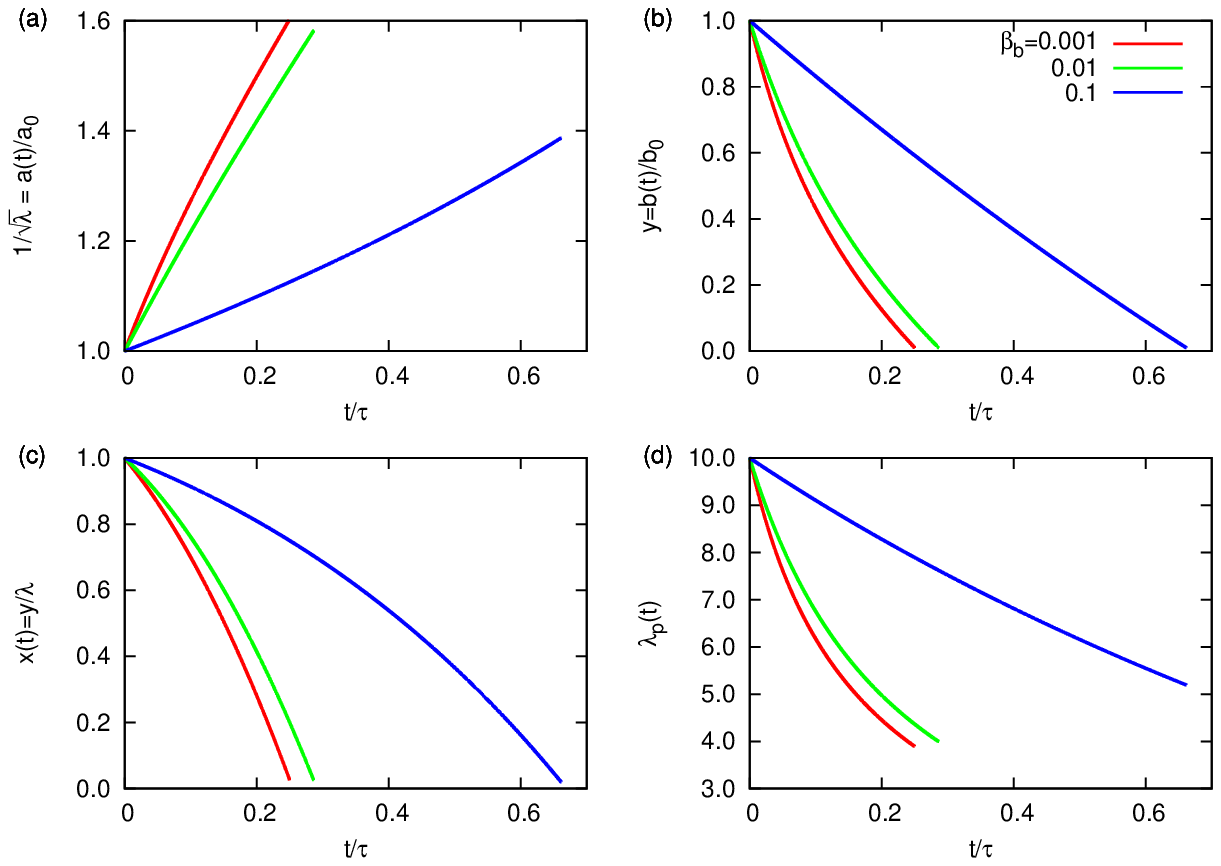}
  \caption{Evolution of the system with $\eta_s/\eta_p=0.01$, $\alpha=100$, $\beta_{\rm t}=0.1$, and $\lambda_{p0}=10$. (a) The radius of the string $a(t)=a_0/\sqrt{\lambda}$. (b) The length of the string $b(t)=yb_0$. (c) The fraction of the string material $x(t)=y/\lambda$. (d) The extension of the polymer chain $\lambda_p(t)$. The time in the horizontal axis is scaled by the relaxation time $\tau$.} 
  \label{fig:BSB1}
\end{figure}

Figure~\ref{fig:BSB1} shows the numerical results for $\eta_s/\eta_p=0.01$, $\alpha=100$, $\beta_{\rm t}=0.1$, $\lambda_{p0}=10$, and $\beta_{\rm b}=0.001, 0.01, 0.1$.   
For all cases, the radius of the string increases with time [Fig.~\ref{fig:BSB1}(a)] and the length of the string decreases with time [Fig.~\ref{fig:BSB1}(b)].
The volume of the filament, on the other hand, decreases in time [Fig.~\ref{fig:BSB1}(c)], so the material flows from the filament to the beads. 

\vspace{0.2cm}

\subsection{finite $\beta_{\rm b}$ case}

We have shown that  when $\beta_{\rm b} \gg \beta_{\rm t}$, the filament gets thinner (Sec. \ref{sec:CaBER}), and when $\beta_{\rm b} \ll \beta_{\rm t}$, the filament gets thicker.
We now consider the situation that $\beta_{\rm b}$ is comparable with $\beta_{\rm t}$. 

\begin{figure}[tbp]
  \centering
  \includegraphics[width=1.0\textwidth]{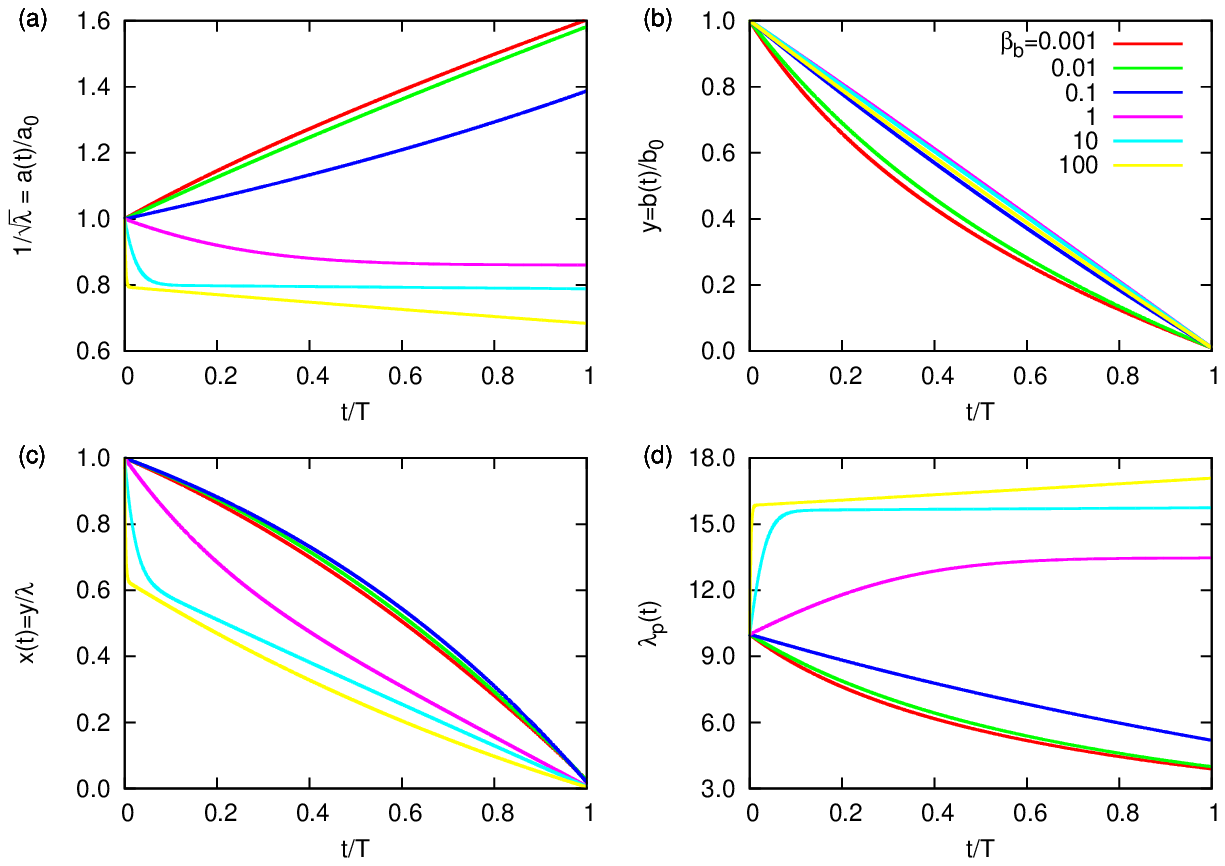}
  \caption{Evolution of the system with $\eta_s/\eta_p=0.01$, $\alpha=100$, $\beta_{\rm t}=0.1$, and $\lambda_{p0}=10$. (a) The radius of the string $a(t)=a_0/\sqrt{\lambda}$. (b) The length of the string $b(t)=yb_0$. (c) The fraction of the string material $x(t)=y/\lambda$. (d) The extension of the polymer chain $\lambda_p(t)$. The time in the horizontal axis is scaled by the touching time $T$.}
  \label{fig:BSB2}
\end{figure}

Figure~\ref{fig:BSB2} shows the numerical results when $\beta_{\rm b}$ is increased from 0.001 to 100 (with $\beta_{\rm t}=0.1$).  
The other parameters are the same as those in the previous section.
Since the dynamics of the filament takes place at different time scale, we plot the time evolution with the time normalized by the touching time $T$ at which the length of the filament becomes zero. 
We estimate the touching time by the time when $y=0.01$.    
It can be seen that when $\beta_{\rm b}$ is increased, the filament radius changes from increasing to decreasing [Fig.~\ref{fig:BSB2}(a)]. 

This behavior is understood from Eq.~(\ref{eq:dlam2}).
We consider the situation that the major driving force for the relaxation is the interfacial energy, \emph{i.e.}, the case $\alpha \sqrt{\lambda} \gg \ \lambda_p^2$.  
\begin{eqnarray}
  \dot{\lambda} &=& - \frac{\sqrt{\lambda}}{\tau \beta_{\rm t} y } 
     \left(\frac{1}{2} \lambda_p^2 - \alpha \sqrt{\lambda} \right) 
     - \frac{1}{ \tau \beta_{\rm b} } 
       \left( \lambda_p^2 + \alpha \sqrt{\lambda} \right)  \nonumber \\
  & \simeq &  \frac{\alpha \lambda}{\tau \beta_{\rm t} y } 
       -  \frac{\alpha \sqrt{\lambda}}{ \tau \beta_{\rm b} } .
  \label{eq:dlamB}
\end{eqnarray}
In this case, the first term on the right hand side of Eq.~(\ref{eq:dlamB}) is positive while the second term is negative.  
The first term represents the effect that if the material moves out from the filament region to bead region, the interfacial area (and therefore the interfacial energy) decreases.  
This process decreases the filament radius, and the rate of the process is determined by $\beta_{\rm t}$, the friction constant associated with the flux from filament to beads. 
On the other hand, the second term represents the effect that if the aspect ratio of the filament decreases, the interfacial energy decreases.  
This process increases the filament radius, and the rate of the process is determined by $\beta_{\rm b}$, 
the friction constant associated with the the motion of the beads. 
Therefore if $\beta_{\rm b}$ is small, the second term dominates, and the filament gets thinner with time. 
On the other hand, if $\beta_{\rm b}$ is large, the first term dominates, and the filament gets thicker with time.  

Alternatively, the behavior can be directly explained by the Onsager principle. 
The surface area of the filament $S= 4 \pi a b$ is written in terms of the volume of the filament $V=2\pi a^2 b$ and the length of the filament $b$ as $S=\sqrt{8 \pi} (Vb)^{1/2}$.  
This can be decreased by reducing $V$ keeping $b$ constant, or by reducing $b$ keeping $V$ constant.  The former process involves the fluid transport from filament region to bead region and is 
governed by $\beta_{\rm t}$.  
The latter process involves the motion of the bead, and is governed by  $\beta_{\rm b}$.  
If  $\beta_{\rm t} \ll \beta_{\rm b} $, the former process dominates and the filament get thinner.  
On the other hand, if $\beta_{\rm t} \gg \beta_{\rm b} $, the latter process dominates and the filament get thicker.

\begin{figure}[htbp]
  \centering
  \includegraphics[width=0.6\textwidth]{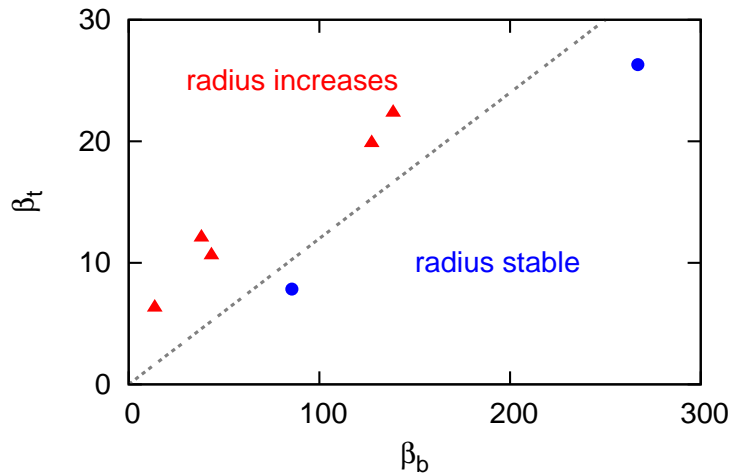}
  \caption{Comparison between the experimental results from Ref. \cite{YangJianmao2008} and the theoretical prediction. The detailed calculation of the data points is described in Appendix \ref{app}. }
  \label{fig:exp}
\end{figure}

This result is in qualitative agreement with the experiments of Yang and Xu \cite{YangJianmao2008}.
They observed that when the initial radius is less than about 16 $\mu$m, the filament radius 
remains constant while the two beads approach each other. 
When the initial radius is larger than 16 $\mu$m, the radius will expand with time.
We plot the $\beta_{\rm b}$ and $\beta_{\rm t}$ values of reported cases in Fig.~\ref{fig:exp}. 
It seems that the two scenario can be separated by the line $\beta_{\rm t} = 0.12 \beta_{\rm b}$. 

\section{Summary}
\label{sec:summary}

In this paper, we have constructed a theoretical model for the time evolution of a viscoelastic filament connecting two mobile droplets.  
We have extended our previous theory for CaBER geometry, and derived time
evolution equations for the filament radius and length. 
In this extended theory, the energy dissipation taking place at the interface between the
filament region and the droplet region is explicitly taken into account. 
We have shown that in the CaBER geometry, the interfacial energy dissipation does not alter the previous results significantly,  while in the present geometry, the interfacial energy dissipation plays a crucial  role. 
This theory explains the experimental results of Yang and Xu \cite{YangJianmao2008} qualitatively.
 
\begin{acknowledgments}
This work was supported by the National Natural Science Foundation of China (NSFC) through the Grant No. 21774004. 
M.D. acknowledges the financial support of the Chinese Central Government in the Thousand Talents Program. 
\end{acknowledgments}

\appendix

\section{Experimental data from Ref.~\cite{YangJianmao2008}}
\label{app}

Here we describe the calculation of $\beta_{\rm b}$ and $\beta_{\rm t}$ from the experimental data in Ref.~\cite{YangJianmao2008}. 
The polymer solution is HPIB 1\% ($\eta_s=60.5$ ${\rm Pa}\cdot{\rm s}$, $\eta_p = 287$ ${\rm Pa}\cdot{\rm s}$), and the medium is PDMS ($\eta_m=12.0$ ${\rm Pa}\cdot{\rm s}$) 
The shear viscosities are measured at 22$^{\circ}$C (Table I from Ref.~\cite{YangJianmao2008}), so we only used the experimental data performed at 19$^{\circ}$C and 30$^{\circ}$. 
There are total 7 experiments (Table II from Ref.~\cite{YangJianmao2008}), among which 2 cases the filament is stable and 5 cases the filament increases with time. 
In experiments, the two droplets do not have the same size, so we use the averaged value in $2R_0$. 
The value of $\beta_{\rm b}$ and $\beta_{\rm t}$ are calculated using Eqs.~(\ref{eq:betab}) and (\ref{eq:betat}).

\begin{table}
\begin{tabular}{ p{1cm} p{2cm} | c c c | c c }
\hline
\hline
No. &   & $\quad$ $2 a_0$ [$\mu$m] $\quad$ & $\quad$ $2 b_0$ [$\mu$m] $\quad$ & $\quad$ $2R_0$ [$\mu$m] $\quad$ & $\quad$ $\beta_{\rm b}$ (\ref{eq:betab}) $\quad$ & $\quad$ $\beta_{\rm t}$ (\ref{eq:betat}) $\quad$ \\
\hline 
(1)  & stable &    10.6 & 198 & 558   & 85.5 & 7.85 \\
(2)  & stable &    11   & 689 & 539   & 267  & 26.3 \\
(6)  & increase &  17   & 803 & 527   & 127  & 19.8 \\
(7)  & increase &  18   & 958 & 539   & 139  & 22.4 \\
(8)  & increase &  26.4 & 667 & 520.5 & 43.3 & 10.6 \\
(9)  & increase &  34   & 977 & 518.5 & 38.1 & 12.1 \\
(11) & increase &  49   & 736 & 511   & 13.6 & 6.31 \\
\hline
\end{tabular}
\caption{Calculation of $\beta_{\rm b}$ and $\beta_{\rm t}$ from Ref.~\cite{YangJianmao2008}.}
\end{table}

\newpage
\bibliography{ve}


\end{document}